\documentclass[12pt,preprint]{aastex}

\newcommand{\bm}[1]{\mbox{\boldmath $#1$}}
\newcommand{\daw}{{\rm daw}}
\newcommand{\TA}{T_{\rm A}}
\newcommand{\tA}{t_{\rm A}}
\newcommand{\VA}{V_{\rm A}}
\newcommand{\vA}{v_{\rm A}}

\slugcomment{Accepted for publication in ApJ 04/14/2004}

\shorttitle{Stability of Magnetic Reconnection}
\shortauthors{Hirose et al.}

\begin{document}

\title{NUMERICAL EXAMINATION OF THE STABILITY OF
AN EXACT TWO-DIMENSIONAL SOLUTION 
FOR FLUX PILE-UP MAGNETIC RECONNECTION}

\author{Shigenobu Hirose}
\affil{Department of Physics and Astronomy, Johns Hopkins University,
    Baltimore, MD 21218-2686, USA}
\email{shirose@pha.jhu.edu}
\author{Yuri E. Litvinenko}
\affil{Institute for the Study of Earth, Oceans, and Space,
   University of New Hampshire, Durham, NH 03824-3525, USA}
\email{yuri.litvinenko@unh.edu}
\author{Syuniti Tanuma; Kazunari Shibata}
\affil{Kwasan and Hida Observatories, Kyoto University, 
Yamashina-ku, Kyoto 607-8471, Japan}
\email{tanuma@kwasan.kyoto-u.ac.jp; shibata@kwasan.kyoto-u.ac.jp}
\author{Masaaki Takahashi}
\affil{Department of Physics and Astronomy, Aichi University of Education,
Kariya, Aichi 448-8542, Japan}
\email{takahasi@phyas.aichi-edu.ac.jp}
\author{Takayuki Tanigawa}
\affil{Academia Sinica, Institute of Astronomy and Astrophysics,
Taipei 106, Taiwan, R.O.C.}
\email{tanigawa@asiaa.sinica.edu.tw}
\author{Takahiro Sasaqui}
\affil{Department of Astronomy, University of Tokyo,
Bunkyo-ku, Tokyo 113-0033, Japan}
\email{sasaqui@th.nao.ac.jp}
\author{Ayato Noro}
\affil{Department of Physics, Chiba University,
Inage-ku, Chiba 263-8522, Japan}
\email{noro@astro.s.chiba-u.ac.jp}
\author{Kazuhiro Uehara}
\affil{Department of Astronomy, Kyoto University,
Sakyo-ku, Kyoto 606-8502, Japan}
\email{uehara@kwasan.kyoto-u.ac.jp}
\author{Kunio Takahashi}
\affil{Department of Earth Science, Ibaraki University,
Mito, Ibaraki 310-8512, Japan}
\email{kutaka@env.sci.ibaraki.ac.jp}
\author{Takashi Taniguchi}
\affil{Department of Physics, Tokyo University of Science, 
Shinjuku, Tokyo 162-8601, Japan}
\email{taka0825-0612@mail.goo.ne.jp}
\and
\author{Yuliya A. Terekhova}
\affil{Institute for the Study of Earth, Oceans, and Space,
   University of New Hampshire, Durham, NH 03824-3525, USA}
\email{yuliya.terekhova@unh.edu}

\begin{abstract}
The Kelvin--Helmholtz (KH) and tearing instabilities are likely to be important for the process of fast magnetic reconnection that is believed to explain the observed explosive energy release in solar flares. Theoretical studies of the instabilities, however, typically invoke simplified initial magnetic and velocity fields that are not solutions of the governing magnetohydrodynamic (MHD) equations. In the present study, the stability of a reconnecting current sheet is examined using a class of exact global MHD solutions for steady state incompressible magnetic reconnection, discovered by Craig \& Henton. Numerical simulation indicates that the outflow solutions where the current sheet is formed by strong shearing flows are subject to the KH instability. The inflow solutions where the current sheet is formed by a fast and weakly sheared inflow are shown to be tearing unstable. Although the observed instability of the solutions can be interpreted qualitatively by applying standard linear results for the KH and tearing instabilities, the magnetic field and plasma flow, specified by the Craig--Henton solution, lead to the stabilization of the current sheet in some cases. The sensitivity of the instability growth rate to the global geometry of magnetic reconnection may help in solving the trigger problem in solar flare research.
\end{abstract}


\section{INTRODUCTION}

The current consensus is that the explosive release of magnetic 
energy by virtue of magnetic reconnection in the corona is 
responsible for the observed solar flares 
\citep[e.g.][]{Tsuneta(1996),Shibata(1999),
Priest & Forbes(2000),Moore et al.(2001)}. 
Reconnecting current sheets in the solar corona are believed 
to form and evolve on the scale of days and weeks, 
allowing the free magnetic energy in the corona to be 
accumulated in the sheet. Eventually the sheet 
is disrupted by an instability, leading to the flare energy release on 
the scale of minutes and hours. 

A well-known difficulty in the 
flare theory is the so-called trigger problem, which relates 
to the fact that a relatively slow evolution of the sheet, 
described by ideal magnetohydrodynamics (MHD), 
is followed by rapid destabilization when resistive effects 
in the sheet become important 
\citep[e.g.][]{Wang & Bhattacharjee(1994),Hirose et al.(2001)}. 
The Kelvin--Helmholtz (KH) and tearing instabilities are typically 
invoked to interpret the transition to the latter stage of the evolution 
\citep[e.g.][for a review]{Knoll & Chacon(2002),Biskamp(1994)}. 

Given possible applications to the problem of current sheet 
evolution and disruption in the solar corona,
it is interesting to determine whether the standard results 
on the KH and tearing instabilities are robust with respect to 
the influence of the reconnection flow in the sheet and the global 
geometry of the reconnection solution. This provides the motivation 
for the present work.

A weakness of most previous analyses of the current sheet 
instabilities is that they assumed an initial unperturbed state 
that was not a solution to the resistive steady MHD equations.  
Previous studies typically assumed a ``quasi-equilibrium'' 
one-dimensional $\tanh$-like profile of magnetic field 
either with zero velocity 
\citep{Malara et al.(1991),Malara et al.(1992),Dahlburg et al.(1992)}
or with a one-dimensional velocity proportional to 
the magnetic field \citep{Dahlburg & Einaudi(2001)}.
Effects of the plasma flow and the resistive diffusion  
were not exactly balanced for these initial conditions 
unless a nonuniform electric resistivity in the sheet was 
postulated \citep{Schumacher & Seehafer(2000)}. 
Left alone, such current sheet would resistively decay.
One might also argue that heuristic one-dimensional models 
for the unperturbed magnetic and velocity field could only 
describe the immediate vicinity of the current sheet rather than 
a global MHD solution for magnetic reconnection. 
We are aware of one work in which the unperturbed state 
is an exact solution of MHD equations with a uniform and 
constant electric resistivity: \citet{Phan & Sonnerup(1991)} 
investigated linear stability against tearing mode of 
a one-dimensional current sheet supported by a 
stagnation-point flow in the plane perpendicular to 
the magnetic field vector.

In this paper, we use MHD simulations to examine 
the stability of a reconnecting current sheet described by an exact 
two-dimensional solution for magnetic reconnection 
\citep{Craig & Henton(1995)}. 
The Craig--Henton solution provides analytical description of 
flux pile-up magnetic reconnection in incompressible plasma 
of arbitrary uniform electric resistivity. The resistive diffusion is exactly 
balanced by a sheared stagnation-point flow, so that 
steady-state MHD equations are satisfied globally. 
Notably, we limit consideration to two-dimensional 
instabilities of the Craig--Henton solution, 
although previous studies indicated that 
three-dimensional ideal instabilities of two-dimensional 
current sheets can grow much faster than any 
two-dimensional instabilities 
\citep[e.g.][]{Schumacher & Seehafer(2000),Dahlburg et al.(1992)}.

Recall that flux pile-up current sheets have been repeatedly 
studied and generalized not only in two but also in three dimensions
\citep[e.g.][]{Craig et al.(1995),Litvinenko & Craig(2000)}. 
It is understood that both the Craig--Henton solution 
and those considered by \citet{Sonnerup & Priest(1975)} 
and \citet{Phan & Sonnerup(1991)} are 
particular cases of a general family of solutions 
for flux pile-up merging and reconnection.
The analytical nature of the solutions made it possible 
to demonstrate explicitly many of the characteristics of 
magnetic reconnection, such as the appearance of small
 length scales defined by resistivity, magnetic sling shots, 
and Alfv\'enic outflows
\citep[e.g.][]{Litvinenko & Craig(1999),Litvinenko & Craig(2000)}. 
It is worth stressing that, 
although the pile-up reconnection solution is locally equivalent 
to the Sweet--Parker current sheet as far as reconnection 
scalings with resistivity are concerned, the two models are not
identical globally. The Sweet--Parker model 
postulates a uniform advection region outside the current sheet,
whereas the Craig--Henton solution explicitly describes 
the magnetic field amplification caused by the non-uniform
velocity field outside the sheet.

The paper is organized as follows.
In Section 2, we briefly review the steady-state flux pile-up 
reconnection solutions in two dimensions and identify two types 
of the solutions depending on the sign of a plasma velocity 
component at the boundary. In Section 3, 
we present the numerical methods for examining the 
stability of the solutions and describe the numerical results. 
We compare the results with predictions of the linear theory of 
the KH and tearing instabilities, discuss the limitations of the local 
approach and suggest possible solar applications in Section 4. 
We summarize in Section 5.

\section{REVIEW OF THE EXACT FLUX PILE-UP RECONNECTION SOLUTION}

We begin by reviewing the Craig--Henton solution that satisfies 
the steady-state equations of incompressible non-viscous resistive 
MHD in a planar geometry. Using the Poisson bracket
$[\psi,\phi]\equiv\psi_x\phi_y-\psi_y\phi_x$, the non-dimensionalized 
basic equations can be written as
\begin{eqnarray}
&&[\nabla^2\phi,\phi]=[\nabla^2\psi,\psi], \\
&&E+[\psi,\phi]=\eta\nabla^2\psi,
\end{eqnarray}
where $\psi$ and $\phi$ are, respectively, a flux function for 
the magnetic field $\bm{B}=\nabla\times\psi\hat{\bm{z}}$ and a stream
function for the velocity field ${\bm{v}}=\nabla\times\phi\hat{\bm{z}}$. 
The problem is specified by two dimensionless parameters 
$E$ and $\eta$. The uniform electric field $E$, normalized by 
$\vA B_0$ (in electromagnetic units), characterizes the reconnection rate. 
The inverse Lundquist number $\eta = \eta_0 / (\vA L)$ is based on a 
uniform magnetic diffusivity $\eta_0$. Here and in what follows, 
$B_0$, $\rho_0$, and $L$ are the reference values of the magnetic 
field, plasma density, and spatial length scale, 
$\vA = B_0 / \sqrt{4\pi \rho_0}$ is the reference Alfv\'en speed.

An exact reconnection solution satisfying the equations above is given by 
\begin{eqnarray}
&&\bm{B}(x,y)=\left(\beta x,-\beta y\right)+\left(0,-\frac{E}{\eta\mu} \daw\left(\mu x\right)\right),\label{eq:Bsol}\\
&&\bm{v}(x,y)=\left(\alpha x,-\alpha y\right)+\left(0,-\frac{\beta}{\alpha}\frac{E}{\eta\mu} \daw\left(\mu x\right)\right),\label{eq:vsol}
\end{eqnarray}
where $\daw(x)\equiv\int_0^x\exp\left(t^2-x^2\right)dt$ is the Dawson
function, and the dimensionless parameters $\alpha$ and $\beta$ define, 
respectively, the amplitude and the shear of 
the velocity field \citep{Craig & Henton(1995)}. 
These parameters are used to define $\mu$ according to
\begin{equation}
\mu^2={\beta^2-\alpha^2 \over 2\alpha\eta}.
\end{equation}
Clearly $\alpha$ and $\beta$ should satisfy 
$(\beta^2-\alpha^2)/\alpha>0$ for the physically interesting case 
of a localized current sheet, in which case the sheet thickness 
scales as $\mu^{-1}$.

The distribution of plasma pressure, 
\begin{equation}
p(x,y)=p_0-\frac{1}{2}\left(\alpha^2\left(x^2+y^2\right)+\left(\frac{E}{\eta\mu} \daw\left(\mu x\right)\right)^2\right)-\beta y\frac{E}{\eta\mu} \daw\left(\mu x\right),\label{eq:psol}
\end{equation}
is derived from the original momentum equation 
\begin{equation}
\left(\bm{v}\cdot\nabla\right)\bm{v}=-\nabla^2\psi\nabla\psi-\nabla p,  
\end{equation}
assuming a constant base pressure $p_0$.

Equations (\ref{eq:Bsol}) and (\ref{eq:vsol}) demonstrate that 
both the magnetic field and the velocity field have a 
stagnation-flow component proportional to
$(x,-y)$, and a shear-flow component proportional to $(0,\daw(\mu x))$. 
The shear width scales as $\mu^{-1}$.

Assuming that $\beta \ge 0$,
the solutions are categorized into two groups according to 
the flow direction along the $x$-axis at the external boundary: 
the outflow solutions with $\alpha > 0$ and $\beta>\alpha$ 
and the inflow solutions with $\alpha < 0$ and $|\alpha|>\beta$. 
In the outflow solution the flux pile-up is accomplished mainly 
by the advection of the $x$-component of magnetic field by 
the flow parallel to the $y$-axis.
In the inflow solution $\beta=0$ corresponds to the anti-parallel 
merging with stagnation flow 
\citep{Sonnerup & Priest(1975)}, and $\beta>0$ describes the 
fully two-dimensional solution for flux pile-up reconnection, 
which is also called ``reconnective annihilation'' 
\citep{Priest & Forbes(2000)}.

Viscous effects, ignored in the solution above and hence 
throughout this paper, are likely to be important in the region 
of high shear coinciding with the current sheet.
This is potentially more important for the outflow solution 
that is characterized by a strong velocity shear. 
The Craig--Henton formalism has been 
generalized to include viscosity by 
\citet{Fabling & Craig(1996)}.
The latter solution is more complex mathematically. 
This is why in this paper we concentrate on a simpler non-viscous 
case in which a simple analytical expression for the initial state 
is available. Although viscosity effects are neglected without much 
justification in the outflow case, we wish to stress from the outset 
that the neglect is justified much better in the inflow case when 
the flow is only weakly sheared. As we show in Section 4.4,
it is the inflow case that is of primary interest to us in the solar flare 
context. The inflow solution can be tearing unstable, which 
may explain flare energy release by tearing-induced reconnection.

\section{NUMERICAL EXAMINATION OF THE STABILITY OF THE EXACT SOLUTIONS}

\begin{figure}
\plotone{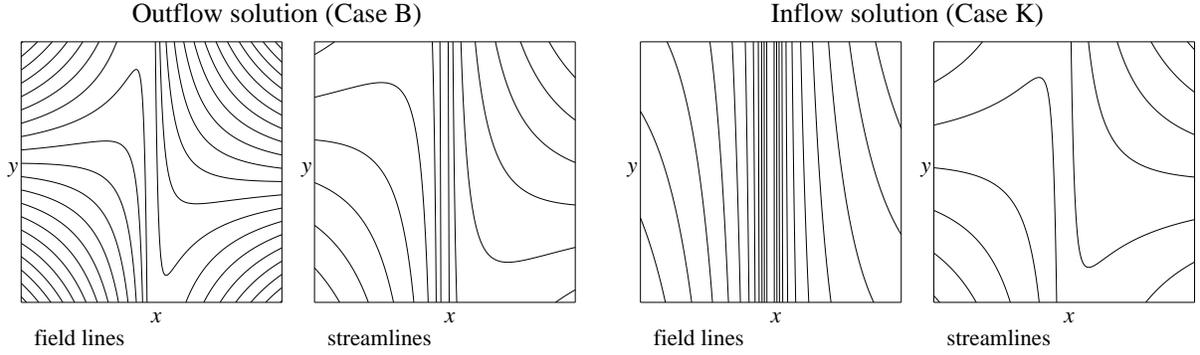}
\caption{Field lines and streamlines of the outflow solution (Case B) 
  and the inflow solution (Case K). The displayed area is 
  $|x|<2, |y|<2$ (the origin is at the center). 
  The densities of field lines and streamlines correspond 
  to $|\bm{B}|$ and $|\bm{v}|$, respectively. Parameters of each case 
  are shown in Table \ref{TAB01}. \label{FIG01}}
\end{figure}

Typical solutions in the outflow case and the inflow case are 
shown in Figure \ref{FIG01}. As noted in the previous section, 
a shear of width $2\mu^{-1}$ is present along the $y$-axis 
in both the magnetic and velocity fields. 
Therefore, we can expect that these solutions can be subject to the 
KH instability and the tearing instability. 
We investigated the stability of the solutions using MHD numerical 
simulations. Specifically we adopted the analytical exact solution as 
the initial condition, perturbed it slightly, and 
examined the time development of the system.

\subsection{Numerical Method}

The basic equations for the numerical simulations are {\it compressible} 
resistive MHD equations in the following dimensionless form:
\begin{eqnarray}
&&\frac{\partial\rho}{\partial t}+\nabla\cdot\left(\rho\bm{v}\right)=0,\\
&&\frac{\partial}{\partial t}\left(\rho\bm{v}\right)+\nabla\cdot\left(\rho\bm{v}\bm{v}-\bm{B}\bm{B}+\left(p+\frac{\bm{B}^2}{2}\right)\mathbf{I}\right)=0,\\
&&\frac{\partial\bm{B}}{\partial t}+\nabla\times\bm{E}=0,\\
&&\frac{\partial}{\partial t}\left(\frac{1}{2}\rho\bm{v}^2+\frac{p}{\gamma-1}+\frac{\bm{B}^2}{2}\right)+\nabla\cdot\left(\left(\frac{1}{2}\rho\bm{v}^2+\frac{\gamma p}{\gamma-1}\right)\bm{v}+\bm{E}\times\bm{B}\right)=0,\\
&& \bm{E}=-\bm{v}\times\bm{B}+\eta\nabla\times\bm{B},
\end{eqnarray}
where $\rho$ is the plasma density 
and $\gamma=5/3$ is the ratio of specific heats. The above equations are 
normalized in the same way as those in \citet{Craig & Henton(1995)} 
and in the previous section, using the reference values of 
length $L$, density $\rho_0$, and Alfv\'en speed ${\vA}$.
The time is normalized with the global Alfv\'en time $\tA\equiv L/\vA$.

To simulate the Craig--Henton solutions derived for {\it incompressible} flow, 
we set the constant base pressure $p_0$ in equation (\ref{eq:psol})
large enough so that the sound speed 
is at least ten times greater than both the fluid speed and the Alfv\'en 
speed. This ensures that the compressible fluid 
behaves almost as the incompressible one. 

The numerical scheme used in the simulations is based on an upwind
scheme with Roe's numerical fluxes \citep{Powell et
al.(1999)}. Second-order accuracy both in space and time is accomplished
by variable extrapolation of Monotone Upstream-centered Schemes for 
Conservation Laws (MUSCL) and multi-step time integration, 
guaranteeing Total Variation Diminishing (TVD) with a non-linear 
limiter function \citep[e.g.][]{Hirsh(1990)}. 
Constraint Transport (CT) scheme is combined with the above scheme 
to develop the magnetic field, ensuring $\nabla\cdot{\bm B}=0$
 \citep{Balsara & Spicer(1999)}. 

The simulation box is set to $|x|<2.5, |y|<2.5$, 
and $500\times500$ constant grids are assigned. This resolution 
is determined so that at least ten grids are assigned in the 
current sheet thickness. The analytical solutions of 
\citet{Craig & Henton(1995)} 
are adopted as the initial conditions for the velocity, 
magnetic field, and pressure. The initial dimensionless density is set to 
unity: $\rho(x,y,0)=1$. In addition, 
a small random velocity perturbation (1\% of the typical 
Alfv\'en velocity $\vA$) is added in the region of $|x|<1, |y|<1$. 
The inverse Lundquist number is uniform and constant: $\eta(x,y,t)=\eta$. 
The boundary values are fixed to the initial values for the whole 
simulation time. The base dimensionless pressure 
$p_0$ is fixed at $5 \cdot 10^4$ in all cases. 
The simulations are performed with the fixed 
Courant number of 0.8, and the corresponding number of time 
steps required for a typical run is of order $10^5$.

As far as the accuracy of the numerical method is concerned, 
we verified that the density fluctuations in the simulation are 
typically of order 0.01\% and never exceed 0.54\%. 
The typical value of $\nabla \cdot {\bf v}$, normalized by 
the ratio of the sound speed and the grid size, is of order 
$10^{-4}$, which is consistent with the magnitude of density 
fluctuations. Therefore the above value for the base pressure 
$p_0$ is indeed large enough to mimic the incompressible behavior.
We also confirmed that the compressible code can maintain the 
unperturbed equilibrium for stable cases. Additionally we 
performed convergence tests for several cases and confirmed 
that the stability does not depend on the resolution. 

\subsection{Overview of the Numerical Results}

Parameters of the solutions of which we examined the stability 
are summarized in Table \ref{TAB01}. In Table \ref{TAB01} and hereafter,
we adopt the Alfv\'en time $\TA\equiv L/\VA$ for the normalization of
time; $\TA$ is {\it defined in each case} based on 
the Alfv\'en speed $\VA$  at the entrance to the current sheet.
This normalization allows us to separate the instability effects under 
study from amplification of magnetic field carried by the reconnection flow. 
The relation between $\TA$ and $\tA = L/\vA$ used in the normalization 
of the basic equations is 
\begin{equation}
\frac{\TA}{\tA}=\frac{\vA}{\VA} \approx \frac{\mu\eta}{E}, \label{eq:tata}
\end{equation}
since $\vA$ is the Alfv\'en speed at the location where the magnitude 
of the magnetic field in equation (\ref{eq:Bsol}) is unity.
Here we have defined $\VA$ ignoring a factor $\approx 0.5$, 
which is the maximum value of the Dawson function, for simplicity.

\begin{figure}
\plotone{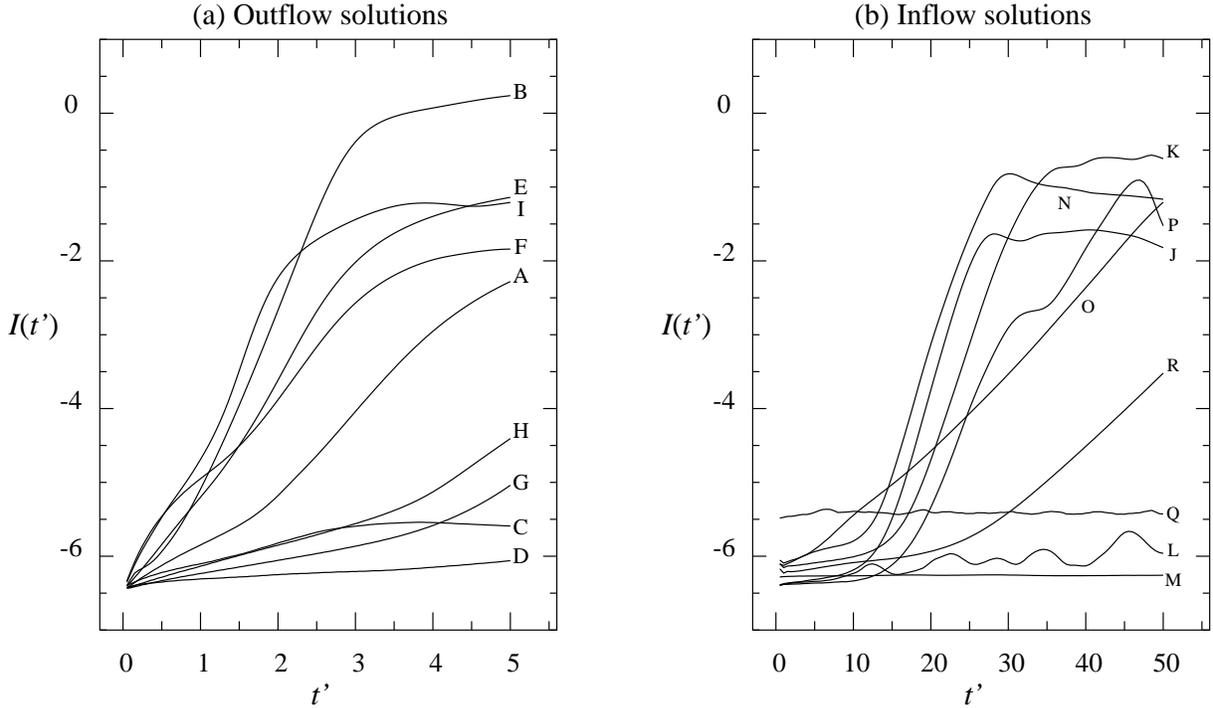}
\caption{The instability growth as measured by $I(t')$ 
(defined in eq. [\ref{eq:grow}]) in outflow solutions (Case A to I) 
and inflow solutions (Case J to R). Note that $t'$ is the time 
normalized with $\TA=L/\VA$. \label{FIG02}}
\end{figure}

The growth of the system in each case is summarized in Figure \ref{FIG02}. 
We quantify the perturbation of the velocity distribution in 
the sheet by plotting the time evolution of 
\begin{equation}
I(t')\equiv\log\left(
\int\!\!\!\int_{|x|<\mu^{-1},|y|<2}|\bm{v}(x,y,t')-\bm{v}(x,y,0)|dxdy \Big/
\int\!\!\!\int_{|x|<\mu^{-1},|y|<2}dxdy\right), \label{eq:grow}
\end{equation} 
where $t'$ is the time normalized by $\TA$.
The duration of the simulation is $0<t'<5$ for the outflow solutions 
and $0<t'<50$ for the inflow solutions. We stopped the simulations 
before the evolution of the system could be affected 
by the fixed boundary conditions.

\begin{figure}
\plotone{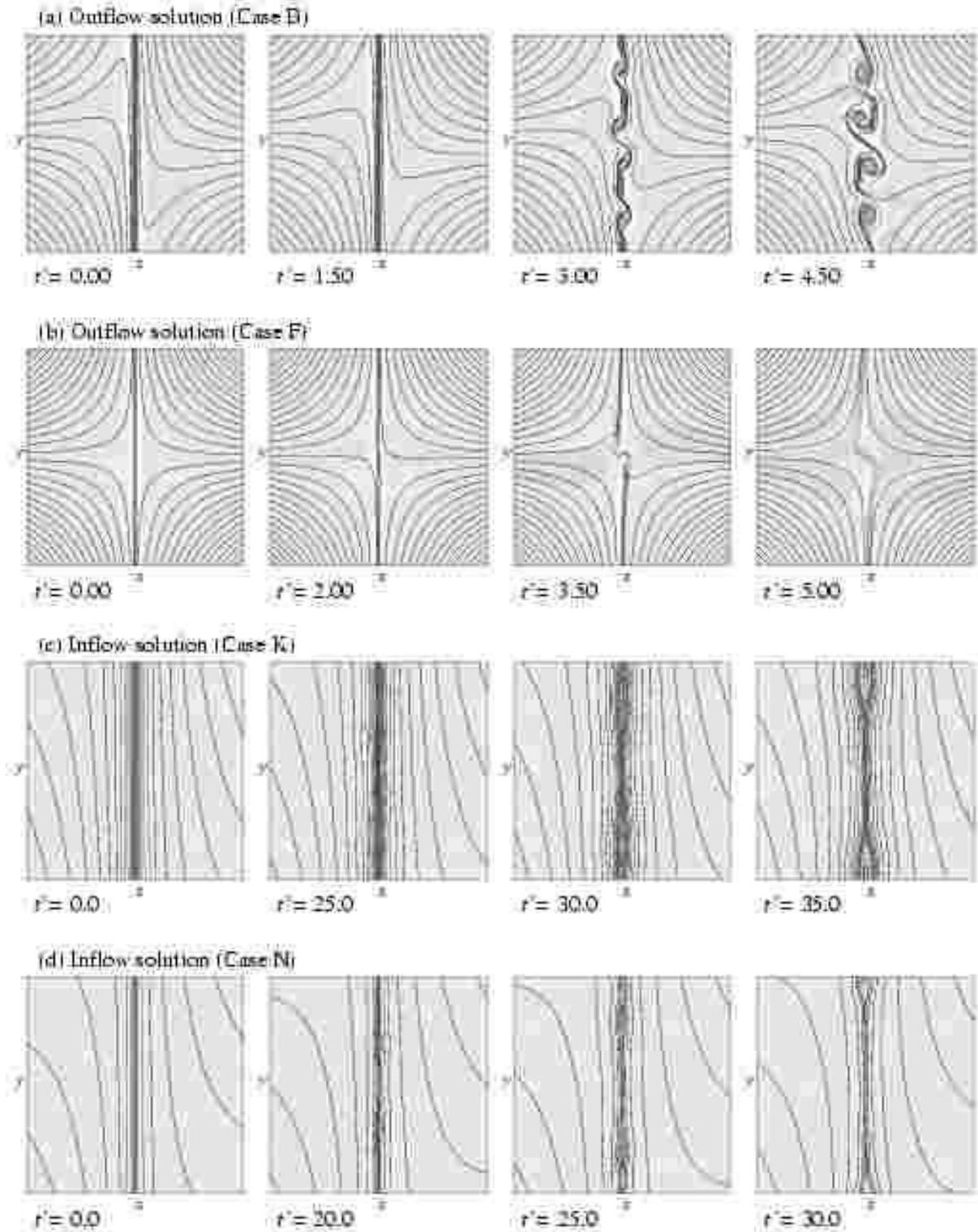}
\caption{Typical time developments of the instabilities 
for outflow solutions (Cases B and F) and inflow solutions (Cases K and N). 
Solid lines represent the magnetic field lines. Gray scale represents 
the $z$-component of the current density; darker gray corresponds to 
a larger current density. The displayed area is 
  $|x|<2, |y|<2$ (the origin is at the center). \label{FIG03}}
\end{figure}

Figure \ref{FIG02} shows that a choice of parameters 
leads to instability of both the outflow and the inflow solutions. 
The nature of the detected instability is, however,  
entirely different for the two types of solutions. 
This can be clearly seen in Figure \ref{FIG03}, in which the 
time evolution of some typical {\it unstable} 
solutions in both cases is shown. In the outflow solutions 
(Fig. \ref{FIG03} (a)(b)), 
the current sheet along the $y$-axis is warped, and then 
vortices with magnetic islands are 
formed, suggesting that the KH instability occurs. 
On the other hand, in the inflow solutions (Fig. \ref{FIG03} (c)(d)), the time 
development of the current distribution is symmetric with respect to the 
$y$-axis, and long and narrow magnetic islands are formed, which 
are characteristic features of the tearing instability.

\begin{figure}
\epsscale{0.7}
\plotone{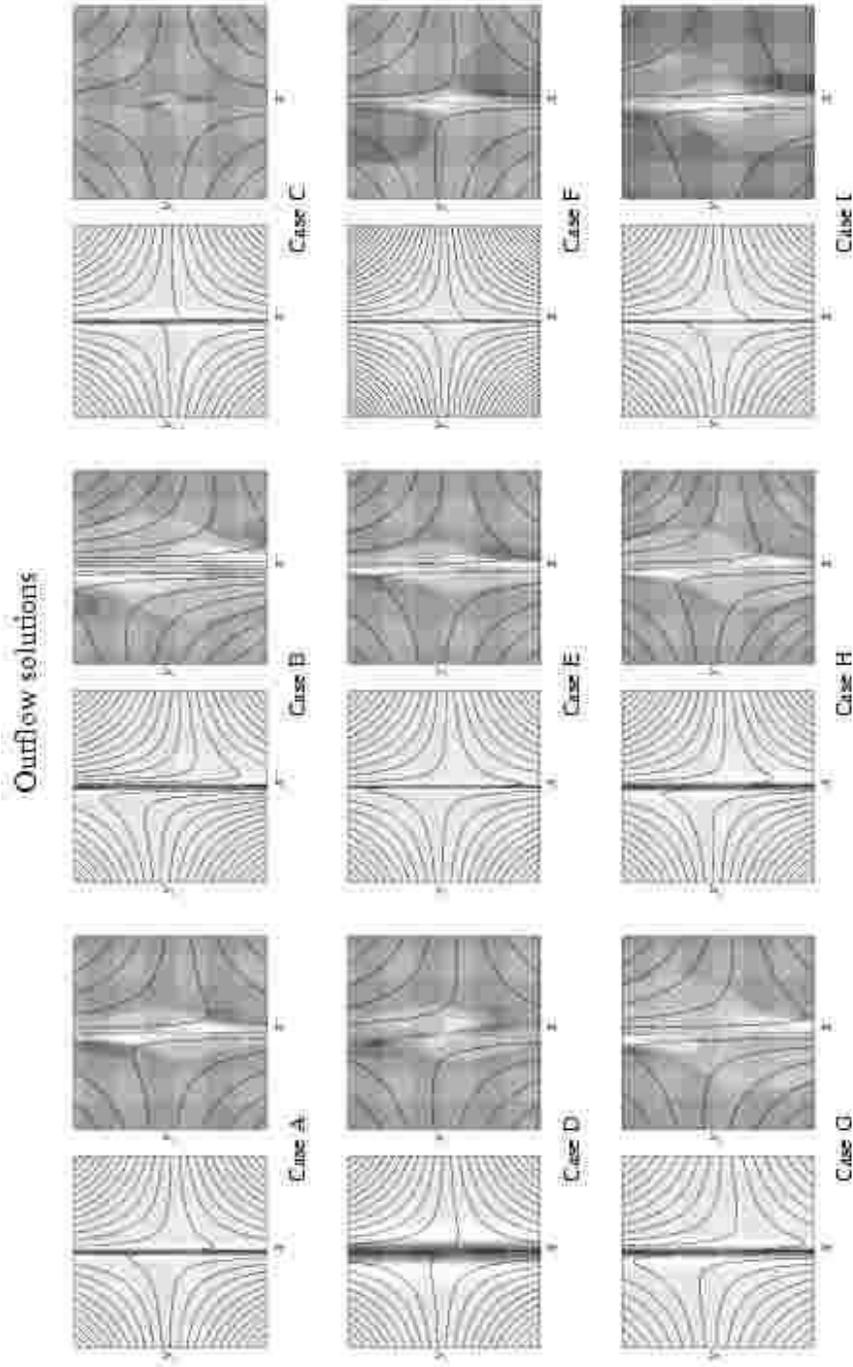}
\caption{Structure of the outflow solutions (Cases A to I). In each 
case, the left figure shows the distribution of the current density 
(gray scale) and field lines (solid lines), while the right 
figure shows the streamlines (solid lines) and $\log_{10}
(E_{\rm K}/E_{\rm M})$ (gray scale), where $E_{\rm K}$ is the kinetic energy 
and $E_{\rm M}$ is the magnetic energy. 
The gray scale in the right figure has six 
steps, $\sim-2$ (lightest), $-2\sim-1$, $-1\sim0$, $0\sim1$, $1\sim2$, 
$2\sim$ (darkest). The displayed area is 
  $|x|<2, |y|<2$ (the origin is at the center).
\label{FIG04}}
\end{figure}

\begin{figure}
\epsscale{0.7}
\plotone{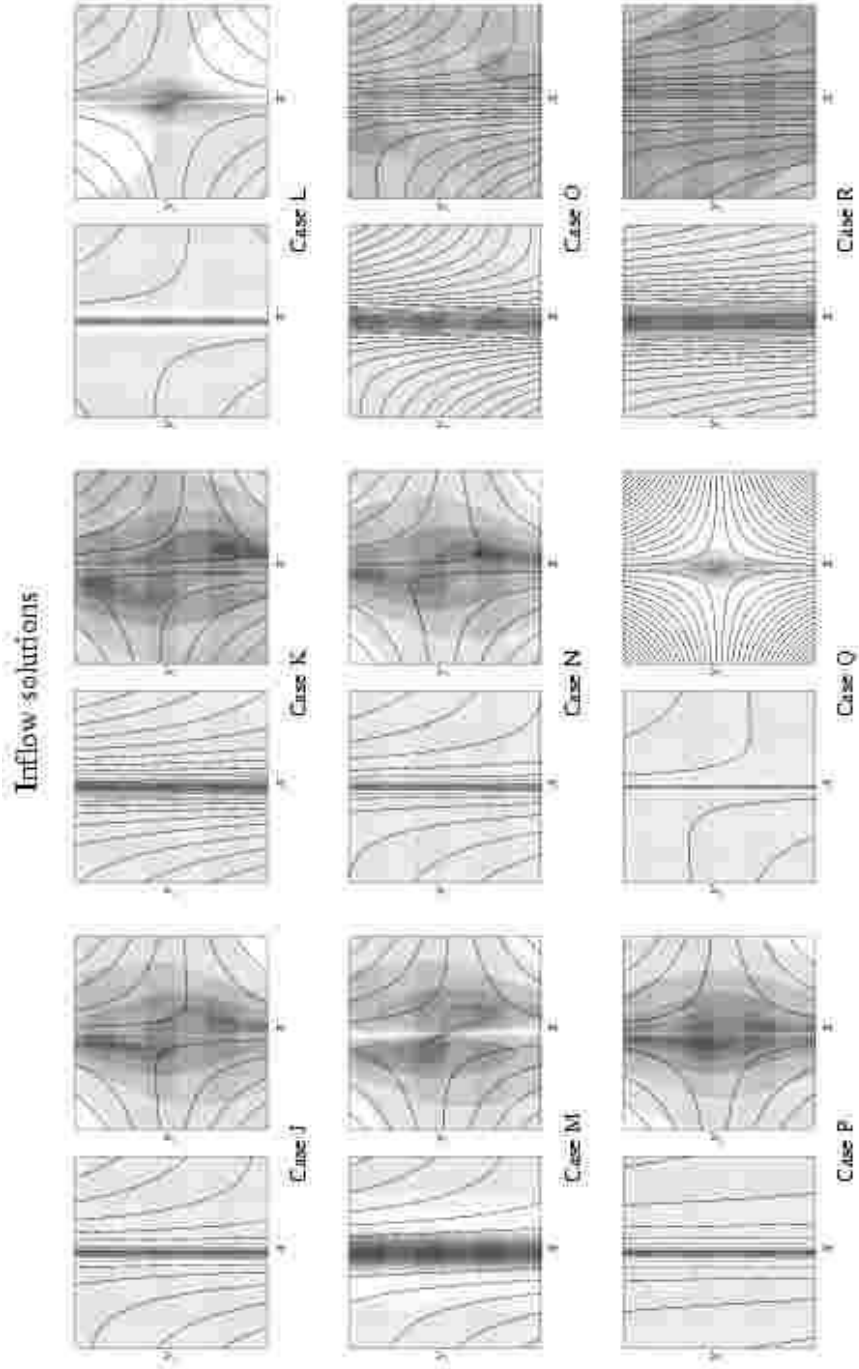}
\caption{Structure of the inflow solutions (Cases J to R). See 
the caption of Figure \ref{FIG04}. \label{FIG05}}
\end{figure}

The difference of the unstable behavior comes from the fact that
the velocity shear is stronger than the magnetic shear 
in the outflow solutions, $\beta>\alpha$, 
and vice versa in the inflow solutions, $\beta<|\alpha|$ 
(see the terms multiplying the Dawson function in 
eqs. [\ref{eq:Bsol}] and [\ref{eq:vsol}]).
These features can be seen in Figures \ref{FIG04} and \ref{FIG05}, 
which show the distributions of the ratio of the kinetic energy 
to the magnetic energy. In the outflow (inflow) solutions, 
the kinetic (magnetic) energy is dominant near the current sheet.

\section{INTERPRETATION OF THE NUMERICAL RESULTS}

In this section, we discuss the parameter dependencies of 
the instabilities on the basis of the standard linear theories 
of the KH instability for the outflow solutions and 
the tearing instability for the inflow solutions. 
We also point out that the standard results, 
derived for uniform initial conditions outside the current sheet, 
have limited applicability to the pile-up solutions, 
characterized by strongly nonuniform profiles of 
the magnetic and velocity fields outside the sheet.

\subsection{KH Instability of the Outflow Solution}

The standard linear growth rate $\omega$ for the KH instability 
is given by the following expression 
\citep{Chandrasekhar(1981), Michael(1955)}:
\begin{equation}
\omega=k\sqrt{\left(\frac{\Delta U}{2}\right)^2-\VA^2},
\end{equation}
where $k$ is the wave number, $\VA$ is the local Alfv\'en 
speed, and $\Delta U$ is the difference in plasma velocities at the 
opposite edges of the sheet. Note that \citet{Chandrasekhar(1981)} 
discussed in detail the case of a uniform magnetic field 
whereas \citet{Michael(1955)} allowed for different 
magnetic field strengths on the two sides of a discontinuity. 
The expressions for $\omega$ are identical, however, if 
the magnetic field is symmetrical on either side of the current sheet, 
which is the case for the Craig--Henton solution.

When applying the formula for the instability growth rate to the 
Craig--Henton solution, it should be remembered that a continuous 
velocity profile has the effect of limiting the KH instability to 
large wave lengths, so that the most unstable mode has $k$ of 
order the inverse of the sheet thickness $l$
\citep{Chandrasekhar(1981)}. 
In this case, since $l=L/\mu$,
\begin{equation}
k_{\rm max}=\mu/L.\label{eq:k-KH}
\end{equation}
Recalling equations (\ref{eq:vsol}) and (\ref{eq:tata}), 
it is straightforward to determine the maximum dimensionless 
linear growth rate: 
\begin{equation}
\omega_{\rm max}t_{\rm A}=(k_{\rm max}L)\sqrt{\left(\frac{\Delta U/2}{\vA}\right)^2-\left(\frac{\VA}{\vA}\right)^2}=\mu\sqrt{\left(\frac{\beta}{\alpha}\frac{E}{\eta\mu}\right)^2-\left(\frac{E}{\eta\mu}\right)^2}=\frac{E}{\eta}\frac{\sqrt{\beta^2-\alpha^2}}{\alpha}.
\end{equation}
The corresponding growth rate 
normalized with $\TA$ (see eq. [\ref{eq:tata}]) is given by
\begin{equation}
\omega_{\rm max}T_{\rm A}=\omega_{\rm max}t_{\rm A}\frac{T_{\rm A}}{t_{\rm A}}=\frac{\beta^2-\alpha^2}{\sqrt{2\alpha^3\eta}}.  \label{eq:KH}
\end{equation}
Linear theory predicts that, 
in order to overcome the stabilizing effect of the magnetic field, 
the shear has to be large enough, $\beta>\alpha$, 
which is the case for the outflow solution. 

A major difference between the geometry of 
the standard KH instability and the outflow Craig--Henton solution is 
the strongly nonuniform profile of the reconnecting magnetic field 
and the associated presence of the current sheet at the surface 
$x=0$. The stabilization condition, however, should not depend 
on whether or not the magnetic field is uniform. In fact 
it is clear on physical grounds that in either case the solution is 
stable as long as the average Alfv\'en speed exceeds the fluid 
velocity difference \citep[e.g.][]{Biskamp(2003)}. On the other 
hand, we will see below that the growth rate of unstable solutions 
can be strongly modified by the global nonuniformity of the field 
outside the current sheet. 

Note that the resistivity in our simulations is small enough for
the time scale of the instability $(\omega_{\rm max})^{-1}$
to be shorter than the diffusion time scale $\tau_{\rm d}$
over the current sheet thickness $l$. The
diffusion time scale is defined as $\tau_{\rm d}=l^2/\eta_0$ or
$\tau_{\rm d}/\TA=E/(\mu^3\eta^2)$ in the dimensionless form. 
The ratio of $\tau_{\rm d}/\TA$ to $(\omega_{\rm max}\TA)^{-1}$ is typically 
several tens, and its smallest value is $5.77$ in Cases C and D.

The dimensionless wave lengths of the fastest growing modes in the 
unstable cases can be explained by the linear theory, 
but the growth rates are smaller by an order of magnitude, compared 
with the predicted values in Table \ref{TAB01}. 
For example, in Case B, the fastest growing mode has 
the wave length of $\approx 0.5$ (see Fig. \ref{FIG03}(a): $t'=3.0$), 
which agrees with the value predicted by the linear theory ($=0.51$), 
while the growth rate ($\approx 2.7$) 
is much smaller than the predicted value ($=21.21$).
One factor that may be responsible for this significant reduction 
in the growth rate is the magnetic field gradient outside the sheet,
which is associated with the reverse electric current in the Craig--Henton
solution. The magnetic field pile-up on either side of the current sheet 
corresponds to the field variation 
of order $\sim E/(\eta\mu)$ over the length scale $\sim \mu^{-1}$ 
(eq. [\ref{eq:psol}]). Since the corresponding energy density 
$\sim (E/(\eta\mu))^2$ is comparable to the kinetic energy of 
the shearing motion (eq. [\ref{eq:vsol}]), 
the KH instability may be suppressed by the nonuniform 
magnetic field outside the sheet.  

\subsection{Tearing Instability of the Inflow Solution}

The maximum growth rate of the tearing instability and the corresponding 
wave number are given by the following formulas, 
ignoring factors of order unity  
\citep{Furth Killeen & Rosenbluth(1963)}:
\begin{eqnarray}
&&\omega_{\rm max}=\frac{1}{\sqrt{\tau_{\rm d}\tau_{\rm A}}},\\
&&k_{\rm max}=\left(\frac{\tau_{\rm A}}{\tau_{\rm d}}\right)^{\frac{1}{4}}\frac{1}{l}, 
\end{eqnarray}
where $\tau_{\rm d}=l^2/\eta_0$ and $\tau_{\rm A}=l/\VA$ are 
the diffusion and Alfv\'en time scales in the current sheet of thickness $l$.
Using equations $\eta=\eta_0/(\vA L), l/L=1/\mu$, and (\ref{eq:tata}), 
we have the following expressions for the 
dimensionless linear growth rate and wave number for the 
tearing instability:
\begin{eqnarray}
&&\omega_{\rm max}t_{\rm A}=\sqrt{\frac{\VA}{\vA}\frac{\eta_0}{\vA L}\left(\frac{L}{l}\right)^3}=\sqrt{\frac{E}{\eta\mu}\eta\mu^3}=\sqrt{\frac{\left(\beta^2-\alpha^2\right)E}{2\alpha\eta}}, \label{eq:omega-tear} \\
&&k_{\rm max}L=\left(\frac{\eta_0}{L\vA}\frac{L}{l}\frac{\vA}{\VA}\right)^\frac{1}{4}\frac{L}{l}=\left(\eta\mu\frac{\mu\eta}{E}\right)^\frac{1}{4}\mu=\left(\frac{\left(\beta^2-\alpha^2\right)^3}{8\alpha^3\eta E}\right)^\frac{1}{4}, \label{eq:k-tear}
\end{eqnarray}
where it should be remembered that 
$\alpha < 0$ and $\beta < |\alpha|$.
The corresponding growth rate normalized with $\TA$ is
\begin{equation}
\omega_{\rm max}T_{\rm A}=\omega_{\rm max}t_{\rm A}\frac{T_{\rm A}}{t_{\rm A}}=\frac{\alpha^2-\beta^2}{2 | \alpha | \sqrt{E}}. \label{eq:tear}
\end{equation}
Thus stronger inflows (large $\alpha$) and weaker 
two-dimensionality (small $\beta$) are the factors that 
make the current sheet more unstable with respect to tearing.
Theory demands that $\lambda_{\rm max}/l$ be
greater than $2\pi$ for the instability, which is the case for 
the inflow solutions (see Table \ref{TAB01}).
 
The analytical estimate for $\omega_{\rm max}t_{\rm A}$ ignores 
the effect of the reconnection flow on the instability. It is easy to see on
physical grounds that a strong flow would carry the growing modes 
out of the reconnection region, thus suppressing the instability 
\citep[see][and references therein]{Biskamp(1994)}. 
A rough condition for such stabilization is simply 
$|\alpha| > \omega_{\rm max}t_{\rm A}$, which becomes
\begin{equation}
\frac{|\alpha|}{1-\beta^2/\alpha^2}>\frac{E}{\eta}\label{eq:stabilize}
\end{equation}
for the flux pile-up reconnection solution under consideration. 
For typical simulation parameters, however, this condition is 
difficult to achieve unless $\beta \to |\alpha|$, indicating 
a broad current sheet extending to the external boundaries.

\citet{Chen & Morrison(1990)} and \citet{Ofman et al.(1991)} 
studied the influence of the velocity shear on the tearing instability 
and found that the instability is suppressed when the flow shear is 
larger than the magnetic field shear at the neutral plane. It is worth 
remarking that the flow geometry assumed in our study is different. 
Recall that the inflow Craig--Henton solution is defined by a weak 
shear, specifically $ \beta < | \alpha | $. This condition is required in 
order to have a localized current sheet (see Section 2). 
Another difference with the problem investigated by 
\citet{Chen & Morrison(1990)} is that the 
Craig--Henton solution possesses the reconnection outflow that also 
have a stabilizing effect. There is a formal similarity between 
stabilization by the velocity shear and the reconnection outflow, 
although the flow geometry is different. 
One also should remember that when the flow shear is large, 
the behavior of the reconnecting current sheet is more complex 
because it is no longer easy to distinguish the effects of the tearing and 
KH instabilities, and both effects are generally 
present \citep{Ofman et al.(1991)}.

Both the wave lengths and the growth rates in the unstable cases 
can be basically explained by the linear theory. For example, 
in Case K, the fastest growing mode 
has the wave length of $\approx 2.5$ 
(see Fig. \ref{FIG03}(c): $t'=25.0$ ) 
and the growth rate of $\approx 0.3$, 
which agree fairly well with the values predicted by the 
linear theory ($\lambda_{\rm max}/L=3.51$ and 
$\omega_{\rm max}\TA=0.33$). Furthermore, comparing 
(c) and (d) in Figure \ref{FIG03}, the fastest 
growing mode in Case N has a shorter wave length and a larger 
growth rate than that in Case K, 
which is also consistent with the linear theory (see Table \ref{TAB01}).

\subsection{Effects of Nonuniform Initial Conditions}

Equations (\ref{eq:KH}) and (\ref{eq:tear}) 
predict that the KH instability occurs 
in the outflow solutions ($\beta>\alpha$) and the tearing instability 
occurs the inflow solutions ($\beta<|\alpha|$) for any values of the 
parameters, $\alpha$, $\beta$, $E$, and $\eta$. The 
instabilities in our simulations can be interpreted using the 
standard linear theories.
We found, however, that the Craig--Henton solution is stable 
for some values of the parameters. 
This indicates that the standard linear results, 
based on uniform initial conditions, are of limited use 
when analyzing the stability of global nonuniform MHD solutions. 

In the application of linear theories to the solutions discussed 
above, we used the values of the velocity and magnetic field 
at the entrance to the current sheet. The solutions at hand, however, 
are neither one-dimensional nor uniform. 
We expect that, if the length scale of one-dimensionality is not 
large enough compared with 
the wave length for the most unstable mode, the instabilities 
will be modified owing to the resulting interaction with 
the external fields. 
In Figure \ref{FIG04} (Figure \ref{FIG05}), the dark (light) color 
region outside the current sheet is the region where the KH (tearing) 
instability will be influenced by the external fields, since the magnetic
(kinetic) energy is dominant there. For example, in Case F,  
the kinetic-energy dominant region (light color region) 
is smaller than that in Case B, and thus the KH instability occurs in a 
smaller region compared with Case B (see Fig. \ref{FIG03} (a)(b)). 

It may be argued, therefore, that the instabilities can be 
suppressed when the effects of two-dimensionality and 
nonuniformity, ignored in the standard treatment, 
become important in the initial configuration. 
Equations (\ref{eq:Bsol}) and (\ref{eq:vsol}), 
describing the structure of the Craig--Henton solution, 
can be used to quantify the expected effect. 
The two-dimensional magnetic field in the outflow solution, 
for example, should start influencing the shearing velocity profile 
when the first term of equation (\ref{eq:Bsol}) becomes 
large compared with the second term of equation (\ref{eq:vsol}), 
that is when $\alpha$ increases (Case H), $E$ decreases (Case C), 
or $\eta$ increases (Case D). 
This may explain why cases C (small $E$) and 
D (large $\eta$) in the outflow solution appear to be stable. 
A similar argument can be given for the inflow solution by comparing 
the first term of equation (\ref{eq:vsol}) and the second term 
of equation (\ref{eq:Bsol}). The two-dimensionality should come 
into play when $\alpha$ increases (Case Q), $\eta$ increases (Case M), 
or $E$ decreases (Case L). Perhaps this is why we did not 
detect instability in cases L (small $E$), M (large $\eta$), 
and Q (large $\alpha$) in the inflow solution.

The lack of instability in cases L, M, and Q is also consistent with the
qualitative equation (\ref{eq:stabilize}). Interestingly, it appears that 
this order-of-magnitude stability condition can be made quantitative if 
its left-hand side is multiplied by a factor of order $10^2$. Recall that 
a similarly large factor for the aspect ratio of a dynamic Sweet--Parker
current sheet, stable with respect to tearing, was suggested 
by \citet{Biskamp(1994)}.

\subsection{Tearing in the Solar Corona}

This work is partly motivated by the idea that tearing of 
a large-scale current sheet in the solar corona can lead to 
impulsive energy release in solar flares 
\citep[e.g.][]{Sturrock(1994)}. The flux pile-up 
solution investigated in this paper is a member of the only exact 
analytical MHD family of solutions for magnetic reconnection that 
we are aware of, and analytical scalings based on these 
solutions may be of general interest. One important 
question is whether the growth rate of the tearing mode is sufficiently 
rapid if the realistic electric resistivity in the solar corona is assumed.
Our numerical results are in quantitative agreement with 
the linear theory of tearing instability as long as the stabilizing 
effects of two-dimensional geometry are small enough. 
We now proceed to show that the instability scalings do imply 
the possibility of rapid conversion of magnetic energy in 
a large-scale current sheet in the corona. 

The time-scale of destabilization, caused by tearing of a current sheet, 
is conveniently defined as $\tau = \omega^{-1}_{\rm max}$. 
As discussed above, our simulations confirm that 
the growth rate for the flux pile-up solution $\omega_{\rm max}$ 
can be estimated using equation (\ref{eq:omega-tear}). 
While solar observations put some constraints on the dimensions 
of reconnection regions, reconnection rates, and the electric resistivity, 
the observations are not detailed enough to specify 
the geometry of reconnective flows, given by the parameters 
$\alpha$ and $\beta$. This is why, in order to apply the formula for 
$\omega_{\rm max}$ to unstable current sheets in solar active regions, 
we rewrite equation (\ref{eq:omega-tear}) as follows:
\begin{equation}
  {\tau \over t_{\rm A}} = 
  {S^{1/2} \over {\tilde B}_s^{1/2}} 
  \left( {l \over L} \right)^{3/2} ,
\end{equation}
where $S = \eta^{-1}$ is the Lundquist number, and 
the dimensionless pile-up factor ${\tilde B}_s = V_A / v_A$ 
is simply the magnetic field at the entrance to the sheet 
relative to the reference value $B_0$.
This equation for $\tau$ reduces to the standard result for tearing 
in the Sweet--Parker current sheet, provided the pile-up 
is absent, in which case ${\tilde B}_s = 1$ and $l/L = S^{-1/2}$.

Adopting $L = 10^{9.5}$ cm, $B_0 = 10^2$ G, and
$v_A = 10^9$ cm s$^{-1}$ as the reference values 
in a solar active region, we can estimate $S \approx 10^{14}$ 
based on the classical value of electric resistivity. Models for 
the structure of flaring current sheets suggest that their 
thickness is unlikely to exceed $l \approx 10^5$ cm 
\citep[e.g.][]{LaRosa(1992),Litvinenko & Craig(2000)}, 
which leads to $l/L \approx 10^{-4.5}$. Coronal magnetic 
fields are unlikely to exceed $10^3$ G, hence we should 
assume a modest pile-up factor ${\tilde B}_s \approx10$. 
These numerical values result in the instability time-scale 
of order the Alfv\'en time $t_A \approx 10^{0.5}$ s. 
Although nonlinear effects eventually come into play, 
this estimate for $\tau$ confirms that pile-up current sheets 
can be unstable to tearing and can be destroyed 
quite rapidly in the corona. This conclusion should remain 
qualitatively valid if current-driven instabilities give rise 
to plasma turbulence and enhanced electric resistivity 
in the sheet. Although the sheet should thicken up due to 
increased resistivity, this effect will be partly balanced by 
the corresponding decrease in the effective Lundquist number. 

A relevant question at this point is whether flux pile-up 
reconnection solutions adopted for our simulations are 
possible in the solar corona at all. It might be argued that, 
since the Craig--Henton solution is unstable for a wide range 
of parameters, it is not realized on the Sun. We wish 
to stress, however, that the simulations also identified 
a parameter regime in which the Craig--Henton solution is stable. 
In particular the stability with respect to tearing in cases 
L, M, and Q is consistent with the qualitative stabilization 
condition given by equation (\ref{eq:stabilize}). 
The stability appears to be intimately related to the 
two-dimensionality of the solution. Equation (\ref{eq:stabilize}) 
indicates that instability is easier to achieve when 
$\beta \to 0$, which corresponds to one-dimensional 
merging of magnetic fields. The shear parameter 
$\beta$ would be difficult to determine observationally. 
As in our estimate for $\tau$, however, we can eliminate 
$\beta$ in equation (\ref{eq:stabilize}) by specifying 
the current sheet thickness $l$. In order to give an 
illustrative example, we assume 
$\alpha \approx 1$ and $l \approx 10^6$ cm. 
Then the stabilization condition can be written as 
$E < (l/L)^2 \approx 10^{-7}$, which corresponds to 
$E < 10^{-4}$ V cm${}^{-1}$ in dimensional units. 
We suggest that stable two-dimensional solutions can 
describe relatively thick (large $l$), slowly 
reconnecting (small $E$), stable current sheets 
in the solar corona. 

Finally, based on this sensitivity of our numerical results 
to the global geometry of magnetic reconnection, 
it would be tempting to suggest that changes in the geometry 
of the solution may be responsible for rapid changes in the rate of 
magnetic energy release by magnetic reconnection and thus 
may trigger solar flares. One caveat is that we only performed 
a parametric study of stability of the steady flux pile-up 
solutions. A rigorous approach to the trigger problem would 
require the solution of a much more complicated task: we would 
have to track the dynamical evolution of a current sheet 
with time-dependent parameters that reflect, for example, 
a slow change in boundary conditions at the photosphere 
due to the emergence of magnetic flux and photospheric flows.

\section{DISCUSSION}

Exact MHD solutions are now available, which describe 
genuine flux pile-up magnetic reconnection, 
as opposed to simple one-dimensional merging 
\citep{Craig et al.(1995)}. 
Since these analytical solutions explicitly demonstrate 
many of the properties of fast reconnection, 
their analysis can be a powerful tool 
in the reconnection theory \citep[e.g.][]{Litvinenko & Craig(2000), 
Heerikhuisen Litvinenko & Craig(2002)}. 
Furthermore, the solutions are known quantitatively to describe 
the results of numerical simulations performed using 
a time-dependent code 
\citep{Heerikhuisen Craig & Watson(2000)}. 

We have examined in this paper the stability of an exact 
two-dimensional solution for flux pile-up reconnection 
in incompressible resistive plasma 
\citep{Craig & Henton(1995)}. 
We used numerical simulations to show that 
the solutions of the outflow type can be Kelvin--Helmholtz (KH)
unstable and the solutions of the inflow type can be 
tearing unstable. Notably, we used the steady Craig--Henton 
solution as an initial condition, thus allowing enough time for 
the instability to develop. Our approach should be contrasted 
with that of a time-dependent simulation 
\citep{Heerikhuisen Craig & Watson(2000)} in which 
a localized current sheet was present for a finite 
(perhaps insufficient) time in a doubly periodic geometry. 
Although the latter simulation provided some evidence that 
the sheet could break up into magnetic islands, it was not clear 
whether the tearing instability was responsible for the phenomenon.

In the outflow solutions, where a current sheet is 
formed by strong shearing flows, the KH instability was observed 
to develop. The numerical results can 
be qualitatively explained by the standard linear theory of 
the KH instability (eq. [\ref{eq:KH}]). 
The absolute values of the growth rates, however, are smaller 
than those predicted. This may be due to the strongly 
nonuniform profile of the magnetic field 
at the entrance to the current sheet.
The inflow solutions are characterized by a fast, weakly sheared 
inflow that leads to a strong magnetic field pile-up and the formation 
of a localized current sheet. These are the conditions under which 
the tearing instability is likely to develop. 
The computed growth rates can be explained 
fairly well by the standard linear theory of the tearing 
instability (eq. [\ref{eq:tear}]). 
The instability, however, is suppressed when the effects of 
two-dimensionality are large enough to modify the evolution of 
the flux pile-up current sheets. 

Thus the numerical results often can be interpreted 
by applying standard results for the KH 
and tearing instabilities to the particular geometries of magnetic field 
and plasma flow, specified by the Craig--Henton solution. 
These instabilities cause turbulent and non-steady evolution of the 
current sheet, which is favorable for fast magnetic reconnection 
\citep[e.g.][]{Shibata & Tajima(2001),Tajima & Shibata(1997)} 
or particle acceleration \citep[e.g.][]{Litvinenko(2003), Aschwanden(2002)}.
In some cases, however, the global nature of the solution 
leads to the suppression of the instabilities. 
Of particular interest is the dependence of the growth rate and 
the most unstable wave length on the parameters of the reconnection 
solution: reconnection rate $E$, resistivity $\eta$, reconnection flow 
amplitude $\alpha$ and shear $\beta$. 
Finally, we note that the sensitivity of the results to the nonuniform 
profiles of the magnetic and velocity fields in the reconnection region, 
as well as the effects of two-dimensionality of the solution, 
may be responsible for rapid changes in the rate of 
magnetic energy release by magnetic reconnection, which in turn may 
help in solving the trigger problem in solar flare research.

\acknowledgments
Numerical computations were carried out on VPP5000 
at the Astronomical Data Analysis Center of the National 
Astronomical Observatory, Japan (project ID: yst15a and hst21a), 
which is an inter-university 
research institute of astronomy operated by Ministry of 
Education, Science, Culture, and Sports, 
and on VPP5000 at the Super Computer Center, Nagoya University,
by the joint research program of the Solar-Terrestrial 
Environment Laboratory, Nagoya University.
This study was initiated as a part of the ACT-JST summer school 
for numerical simulation in Astrophysics and Space Plasma.
One of the authors (Y.E.L.) acknowledges fruitful discussions with
I. J. D. Craig and the support by NSF grants ATM-0136718 
and IP-9910067 and NASA grant NAG5-10852.

\begin{deluxetable}{crrrrrrrrrrrr}
\tablecaption{Parameters of the solutions the stability of which is examined. 
  Cases A to I are outflow solutions ($\alpha>0$), and Cases J to R 
  are inflow solutions ($\alpha<0$). The solutions are characterized 
  by the dimensionless electric field $E$, reconnection flow 
  amplitude $\alpha$ and shear $\beta$, and electric resistivity $\eta$. 
  $L$ is the global length scale, $l = L / \mu$ is the current sheet thickness, 
  $T_A$ is the Alfv\'en time based on $L$ and the magnetic field 
  at the entrance to the sheet. 
\label{TAB01}}
\tablewidth{0pt}
\tablehead{
\colhead{Case} & 
\colhead{} & 
\colhead{$E$} & 
\colhead{$\alpha$} &
\colhead{$\beta$} &
\colhead{$\eta$} & 
\colhead{} & 
\colhead{$\mu$} & 
\colhead{$\omega_{\rm num}\TA$\tablenotemark{a}} & 
\colhead{$\omega_{\rm max}\TA$\tablenotemark{b}} & 
\colhead{$\lambda_{\rm max}/L$\tablenotemark{b}} & 
\colhead{$\lambda_{\rm max}/l$\tablenotemark{b}}
}
\startdata
A &                &0.50 &1.0  &2.0  &0.010 &&12.25&1.2 & 21.21&0.51& 6.3\\
B &(large $E$     )&2.00 &1.0  &2.0  &0.010 &&12.25&2.3 & 21.21&0.51& 6.3\\
C &(small $E$     )&0.05 &1.0  &2.0  &0.010 &&12.25&--- & 21.21&0.51& 6.3\\
D &(large $\eta$  )&0.50 &1.0  &2.0  &0.100 && 3.87&--- &  6.71&1.62& 6.3\\
E &(small $\eta$  )&0.50 &1.0  &2.0  &0.003 &&22.36&1.6 & 38.73&0.28& 6.3\\
F &(large $\beta$ )&0.50 &1.0  &3.0  &0.010 &&20.00&1.2 & 56.57&0.31& 6.3\\
G &(small $\beta$ )&0.50 &1.0  &1.5  &0.010 && 7.91&0.7 &  8.84&0.79& 6.3\\
H &(large $\alpha$)&0.50 &1.3  &2.0  &0.010 && 9.42&0.7 & 11.02&0.67& 6.3\\
I &(small $\alpha$)&0.50 &0.5  &2.0  &0.010 &&19.36&2.6 & 75.00&0.32& 6.3\\
  &                &     &     &     &      &&     &    &      &    &    \\
J &                &1.0  &-1.0 &0.25 &0.005 && 9.68&0.32& 0.47 &2.95&28.6\\
K &(large $E$     )&2.0  &-1.0 &0.25 &0.005 && 9.68&0.28& 0.33 &3.51&34.0\\
L &(small $E$     )&0.2  &-1.0 &0.25 &0.005 && 9.68&--- & 1.05 &1.97&19.1\\
M &(large $\eta$  )&1.0  &-1.0 &0.25 &0.100 && 2.17&--- & 0.47 &6.24&13.5\\
N &(small $\eta$  )&1.0  &-1.0 &0.25 &0.001 &&21.65&0.32& 0.47 &1.97&42.7\\
O &(large $\beta$ )&1.0  &-1.0 &0.80 &0.005 && 6.00&0.11& 0.18 &6.05&36.3\\
P &(small $\beta$ )&1.0  &-1.0 &0.05 &0.005 &&10.00&0.24& 0.50 &2.82&28.1\\
Q &(large $\alpha$)&1.0  &-5.0 &0.25 &0.005 &&22.33&--- & 2.49 &0.84&18.8\\
R &(small $\alpha$)&1.0  &-0.4 &0.25 &0.005 && 4.94&0.10& 0.12 &8.10&40.0\\
\enddata
\tablenotetext{a}{The instability growth rates $\omega_{\rm num}$ 
are roughly computed using Fig. \ref{FIG02}.}
\tablenotetext{b}{In the outflow case, the predicted growth rate 
$\omega_{\rm max}$ 
and the wave length for the most unstable mode $\lambda_{\rm max}$ 
are those for the KH instability 
(eqs. [\ref{eq:KH}] and [\ref{eq:k-KH}]), 
while in the inflow case, those for the tearing instability 
(eqs. [\ref{eq:tear}] and [\ref{eq:k-tear}]).}
\end{deluxetable}

\end{document}